\documentclass[12pt]{article}

\usepackage{amsmath,amssymb,amsthm}
\usepackage{color,bm}
\usepackage{graphicx}
\usepackage{setspace}
\usepackage[margin=1in]{geometry}
\usepackage[labelfont={bf,small},labelsep=endash]{caption}

\usepackage{times}

\usepackage[sort&compress,super,comma]{natbib}

\newcommand\grad{{\bm{\nabla}}}
\newcommand\bW{{\mathbf{W}}}
\newcommand\bv{{\mathbf{v}}}
\newcommand\bx{{\mathbf{x}}}
\newcommand\vI{{\mathbf{I}}}
\newcommand\vO{{\mathbf{O}}}
\newcommand\tD{{\mathbf{D}}}
\newcommand\vPhi{{\mathbf{\Phi}}}

\newcommand\gate[1]{{\textsc{#1}}}

\title{Active matter logic for autonomous microfluidics}

\author
{Francis G. Woodhouse$^{1\ast}$ and J\"orn Dunkel$^{2}$\\
\\
\normalsize{$^{1}$Department of Applied Mathematics and Theoretical Physics,}\\
\normalsize{Centre for Mathematical Sciences, University of Cambridge,}\\
\normalsize{Wilberforce Road, Cambridge CB3 0WA, U.K.}\\[.3cm]
\normalsize{$^{2}$Department of Mathematics, Massachusetts Institute of Technology,}\\
\normalsize{77 Massachusetts Avenue, Cambridge MA 02139-4307, U.S.A.}\\
\\
\normalsize{$^\ast$To whom correspondence should be addressed; E-mail: F.G.Woodhouse@damtp.cam.ac.uk.}
}

\date{}

\begin{document}

\maketitle

\doublespacing

{\bf
\noindent
Chemically or optically powered active matter plays an increasingly important role in materials design but its computational potential has yet to be explored systematically. The competition between energy consumption and dissipation imposes stringent physical constraints on the information transport in active flow networks, facilitating global optimization strategies that are not well understood. Here, we combine insights from recent microbial experiments with concepts from lattice-field theory and non-equilibrium statistical mechanics to introduce a generic theoretical framework for active matter logic. Highlighting conceptual differences with classical and quantum computation, we demonstrate how the inherent non-locality of incompressible active flow networks can be utilized to construct universal logical operations, Fredkin gates and memory storage in set--reset latches through the synchronized self-organization of many individual network components. Our work lays the conceptual foundation for developing autonomous microfluidic transport devices driven by bacterial fluids, active liquid crystals or chemically engineered motile colloids.
}

\clearpage
\onehalfspacing

Active materials~\cite{2013Ge,2013Marchetti_Review} powered by light or chemical 
sources offer intriguing technological and biomedical potential, from targeted drug delivery~\cite{Din:2016aa} 
and microscale reactors~\cite{Tompkins25032014,2014Karzbrun} to tissue engineering~\cite{1993Langer} and energy harvesting~\cite{di2010bacterial,Thampie1501854}. An important subgroup of active materials is fluid-based~\cite{2013Marchetti_Review}, encompassing ATP-driven liquid crystals~\cite{2012Sanchez_Nature,2014Keber_Science}, 
bromine-fueled squirmer droplets~\cite{2011Shashi}, Janus particles~\cite{2008Walther_SM,2016Yan_NMat,2016DiLeonardo_NMat}, 
colloidal rollers~\cite{Bricard:2013aa} and microbial suspensions~\cite{2012Sokolov,2016Wioland_NPhys}. These systems are central to current 
microfluidic soft robotics research~\cite{Snezhko:2011aa} owing to their ability to 
self-assemble into complex structures~\cite{2008Walther_SM,2016Yan_NMat,2016DiLeonardo_NMat}, spontaneously create unidirectional  flows~\cite{2016Wioland_RaceTracks} 
and transport microcargos~\cite{Sokolov19012010,di2010bacterial}. While much has been learned about the ordering principles of active 
fluid systems in the last decade~\cite{2013Marchetti_Review}, their intrinsic computational potential has yet to be systematically explored and exploited~\cite{2015Pearce,2016Nicolau_PNAS,2016Wang_EPJST}.

The recent discovery of collective bacterial spin states~\cite{2016Wioland_NPhys} suggests that self-organized 
active flows can be utilized for microfluidic information storage or transport. Moreover, certain classes of
organisms, such as the slime mold \textit{Physarum polycephalum}~\cite{Tero2010_Science,Adamatzky_Book}, use fluid-mediated computation strategies to solve complex optimization problems, but the decentralized algorithms at work have yet to be deciphered.
Microfluidic technology has been successfully employed to perform universal Boolean computation through sub-millimeter bubbles~\cite{Prakash832,Fuerstman828}, enabling the logical control of chemical micro-reactors in lab-on-a-chip devices. However, bubble logic requires an externally applied pressure difference~\cite{Prakash832,Fuerstman828}, analogous to an applied voltage in a conventional computer.
Ferrofluid droplet computation~\cite{2015Katsikis_NPhy} similarly depends on an external rotating magnetic field `clock'.
 By contrast, active liquids  
can flow spontaneously~\cite{2012Sanchez_Nature,2014Keber_Science,2011Shashi,2016Wioland_RaceTracks}  while still undergoing complex global topological interactions~\cite{2016Woodhouse_PNAS}. This makes active fluids a promising candidate for the implementation of autonomous  computation 
schemes to drive microfluidic reaction, mixing and transport devices and uncover algorithmic principles underlying 
decentralized decision-making in \emph{Physarum}~\cite{Tero2010_Science,Adamatzky_Book} and other organisms.

The computational power of any physical  or biological system is limited by design choices~\cite{Sipser_Book} and thermodynamic constraints~\cite{1961Landauer}.  
Classical Turing-type machines~\cite{Sipser_Book}  sequentially perform localized binary operations while being energetically limited through Landauer's principle~\cite{2012Berut_Nature}. 
Quantum computers~\cite{PhysRevLett.75.4714} exploit nonlocal entanglement but typically require very low operating temperatures to suppress decoherence. Neural circuits~\cite{1986Hopfield_Science} 
rely on feedback loops that can be expensive to maintain~\cite{Lestas:2010aa}. DNA computing~\cite{1994Adleman_Science,1996Lipton_Science} exploits parallelism to counter slow processing speeds. 
The active flow networks (AFNs) considered here operate far from thermal equilibrium and  
realize a nonlocal computation approach that functions at room temperature by combining global incompressibility with local energy conversion constraints. The balance of energy uptake and dissipation forces a microbial or ATP-driven fluid to travel at a preferred speed along micro-channels~\cite{2016Wioland_RaceTracks} while fluid incompressibility imposes topological constraints on the flow network dynamics that enable the implementation of logical operations. A compact mathematical description of AFNs is made possible by a recently proposed mapping onto an effective lattice-field theory~\cite{2016Woodhouse_PNAS}. Here, we use this generic framework to construct active matter logic (AML): we implement universal logical operations, reversible gates  and memory storage in set--reset (SR) latches through the synchronized action of many individual AFN components. We also evaluate the 
robustness of AFN-based computation against noise.

\section*{Results}

\subsection*{Input--output active flow network model}

To construct networks capable of logical operations, we first define closed incompressible AFNs as previously introduced~\cite{2016Woodhouse_PNAS} before augmenting with input--output capability. Our mathematical approach towards describing AFNs takes direct guidance from recent experiments~\cite{2016Wioland_RaceTracks} demonstrating that highly concentrated suspensions of \textit{Bacillus subtilis} bacteria spontaneously self-organize into stable unidirectional flows when confined in narrow microfluidic channels (of width less than $50\,\mu$m).

Modelling a network of narrow channels filled with dense active matter, let $\Gamma$ be an oriented graph comprising vertices $V$ and edges $E$; that is, a graph with every edge assigned an arbitrary directionality. Active flows along the edges of $\Gamma$ are then given by the vector $\vPhi = (\phi_e)$ of fluxes $\phi_e \in \mathbb{R}$ along each edge $e \in E$, where $\phi_e > 0$ and $\phi_e < 0$ represent flow with and against the orientation of $e$, respectively.
Now, let $\tD = (D_{ve})$ be the signed incidence matrix of $\Gamma$ such that $D_{ve}$ is $+1$ if $e$ enters $v$, $-1$ if $e$ leaves $v$, and $0$ if $e$ is not incident to $v$.
To model spontaneous activity-driven flow along the edges while respecting incompressibility at every vertex, $\vPhi$ is taken to obey a pseudo-equilibrium model defined by the Hamiltonian
\begin{align}
\label{eq:H0}
H_0 = \lambda \sum_{e \in E} V(\phi_e) + \tfrac{1}{2}\mu \sum_{v \in V} (\tD \cdot \vPhi)_v^2.
\end{align}
The first term, with coupling constant $\lambda$, models spontaneous active flow $\phi_e \rightarrow \pm 1$ through the double-welled potential $V(\phi_e) = -\tfrac{1}{4}\phi_e^4 + \tfrac{1}{6}\phi_e^6$, as in Landau-type models. This is subject to the soft incompressibility constraint imposed by the second term with coupling constant $\mu$ which, provided $\mu \gg \lambda$, requires the net flux $(\tD \cdot \vPhi)_v$ at each vertex $v \in V$ be approximately zero.
The energy~\eqref{eq:H0} then yields discrete minima where each edge is either flowing, $\phi_e = \pm 1$, or in a non-flowing state $\phi_e = 0$ depending on incompressibility-induced topological frustration.
Note that a sixth-order flux potential~$V$ is necessary to avoid an unphysical hidden symmetry of quartics and guarantee physically realistic~\cite{2016Wioland_RaceTracks} discrete minima $\phi_e \in \{-1,0,+1\}$ rather than the continuum of fractional states that would result from the interaction of a quartic potential with the quadratic incompressibility constraint~\cite{2016Woodhouse_PNAS}.
Other polynomial potentials not sharing the hidden symmetry of quartics could be used as well and yield results similar to those discussed below.

Closed AFNs can be expanded to implement 
input--output capability by allowing non-zero flux through special vertices while preserving incompressibility in the bulk, distinct from conventional neural networks~\cite{1986Hopfield_Science,Hopfield_PNAS}.
An input--output AFN has augmented vertex set $V \cup \partial \Gamma$ comprising bulk vertices $V$, which correspond to the vertices of closed AFNs, and boundary vertices $\partial \Gamma$ of degree~$1$ (that is, incident to exactly one edge).
The boundary vertices $\partial\Gamma = \partial\Gamma_\text{in} \cup \partial\Gamma_\text{out}$ function as flux inputs and outputs: input vertices $\partial \Gamma_\text{in}$ are constrained to have net outward flux according to the prescribed binary input vector $\vI = (I_v)$, whereas output vertices $\partial \Gamma_\text{out}$ remain unconstrained; instead, their net flux $O_v = (\tD \cdot \vPhi)_v$ defines the output vector $\vO = (O_v)$.
This is achieved through the boundary energy
\begin{align}
H_{\partial \Gamma} = \tfrac{1}{2}\mu \sum_{v \in \partial \Gamma_\text{in}} [ (\tD \cdot \vPhi)_v + I_v]^2
\end{align}
constraining $(\tD \cdot \vPhi)_v \approx -I_v$ for all $v \in \partial\Gamma_\text{in}$ when $\mu \gg \lambda$,
which is added to the bulk energy $H_0$ in \eqref{eq:H0}.
Finally, we introduce diode edges $E_+ \subseteq E$ permitting flow only in their positive direction $\phi_e > 0$, as can be realized through geometric channel patterning~\cite{2012Denissenko_PNAS}.
In particular, we always connect appropriately oriented edges in $E_+$ to the input and output vertices~$\partial \Gamma$ to prevent spurious backflow into or out of the system. This is accounted for 
through an additional diode energy $H_+$ satisfying $H_+ = \infty$ if $\phi_e < 0$ for any $e \in E_+$ and zero otherwise, giving a total energy $H = H_0 + H_{\partial \Gamma} + H_+$ for an input--output AFN.

When coupled to an environment that acts as a heat bath, AFNs obey dynamics that are closely related to the Toner--Tu model of polar active fluids~\cite{1995Toner_PRL,1998Toner_PRE}.
In the absence of complex correlation statistics~\cite{2006Schneidman}, the flow~$\vPhi$ is taken to obey the Langevin equation~\cite{2016Woodhouse_PNAS}
\begin{align}
d\vPhi = -\grad H dt + \sqrt{2\beta^{-1}} d\bW_t
\label{eq:langevin}
\end{align}
with inverse noise strength $\beta$, where $\bW_t$ is a vector of independent unit-variance Brownian processes, resulting in a Boltzmann stationary distribution $\propto \exp[-\beta H(\vPhi)]$.
Now, in the limit $\mu \rightarrow \infty$, $\vPhi$ is constrained to the submanifold of incompressible and input-respecting flows---that is, $(\tD\cdot\vPhi)_v = 0$ for all $v \in V$ and $(\tD\cdot\vPhi)_v = I_v$ for all $v \in \partial \Gamma_\text{in}$, respectively---on which the components obey
$
d\phi_e = \lambda \phi_e^3 (1-\phi_e^2) dt + \sqrt{2\beta^{-1}} dW_{t,e}
$.
On the other hand, in a narrow channel whose long axis is parallel to~$\hat\bx$, the lowest-order Toner--Tu equation~\cite{1995Toner_PRL,1998Toner_PRE} for the incompressible uniform polarisation field $\bv = v(t) \hat\bx$ reduces to $dv = \sigma v(1-v^2)dt + \sqrt{2\beta^{-1}}dW_t$ for activity strength~$\sigma$.
The usual Toner--Tu term $v(1-v^2)$ is simply the beginning of a phenomenological gradient expansion and serves an identical purpose to our $\phi_e^3(1-\phi_e^2)$ which, as discussed above, derives from a potential which can be of any double-welled form beyond simple quartic.
Therefore, when many such channels are linked together with flux-conserving boundary conditions in a coarse-grained networked form of an incompressible Toner--Tu model~\cite{2015Chen_NJP,2016Chen_NatCommun,Souslov2016}, AFN physics arise as a result of the shared Landau theory roots.

AFNs allow us to construct logical circuits in a conceptually different fashion to classical logic.
Stable states of an AFN, which are local minima of $H$, comprise flowing edges with $|\phi_e| = 1$ and non-flowing edges with $\phi_e = 0$, subject to conservation of flux at every vertex.
When $\Gamma$ is restricted to vertices of degree at most $3$, as we will do here,
the incompressibility constraint then implies that states must have either zero or two flowing edges incident to every bulk vertex.
Stable states therefore comprise vertex-disjoint paths of flowing edges from each active input to distinct outputs and, where possible, closed vertex-disjoint cycles of flowing edges through other internal vertices.
The configuration energy $H$ is then proportional to the (negative of the) number of flowing edges, favouring states with greater total flow.
Because input--output flows are disjoint, they constrain one another's allowable locations according to the global topology of $\Gamma$. Thus the behavior with one input active can be changed globally by activating a second input, suggesting that complex operations can be computed by appropriately designed networks.
(Note that if a more complex form of $V(\phi_e)$ is employed in Eq.~\eqref{eq:H0} which does not respect $V(\pm 1) < V(0)$, the situation reverses and states with lesser flow become favoured, in contrast to the spontaneous flow character of active matter. In such a scenario, topologically-rooted flow interactions are less common and our control is more limited.)
This is particularly true in the zero-noise limit $\beta \rightarrow \infty$ when the only states---the ground states of $H$---are those with the maximum possible number of edges flowing.
It is in this limit that we have the most control in order to create active logic gates, whose ground states at different input choices have output values yielding particular logical operations, as we now show.

\subsection*{Logic gates}

The elementary operations \gate{and} ($\land$) and \gate{or} ($\lor$) can be realized simultaneously as the ground states of a single small active network (Fig.~\ref{fig:and_or_not}a). The network accepts two inputs, $X$ and $Y$, and its two outputs give $X \land Y$ and $X \lor Y$. This is achieved through a single cross-input coupling edge and the insertion of simple edges before one output ($X \lor Y$) to render it energetically favorable to the other ($X \land Y$) when only one input is active.
For example, consider the input state $X = 0$, $Y = 1$. Since there are no loops in this network, any ground state must comprise a single connected active flow from $Y$ to one of the two outputs. In this case, there are only two such flow states: one flowing from $Y$ to the $X \lor Y$ output, and one from $Y$ to the $X \land Y$ output. However, the former of these has $5$ edges in a flowing state whereas the latter has only $3$ edges flowing. Thus the former has lower energy than the latter, implying that for this pair of inputs there is only one ground state, whose outputs correctly read $X \lor Y = 1$ and $X \land Y = 0$.

The operation \gate{not} ($\lnot$) can also be simply realized (Fig.~\ref{fig:and_or_not}b). Unlike \gate{and/or}, since \gate{not} must output $1$ from an input of $0$, flux conservation demands an additional power leg permanently fixed at $1$; conversely, to output $0$ for an input of $1$ with power also present, conservation demands two ground legs, whose value is ignored, down which the input and power can be dumped.
These concepts are familiar from traditional conservative logic~\cite{Fredkin82_IJTP}, wherein logical operations are performed using elements that route unit signals around a circuit while neither creating nor destroying those signals. The  axioms of conservative logic also demand one-to-one mapping of input tuples to output tuples, termed reversibility, for all operations, which is not necessarily implied by conservation. However, simultaneously conservative and reversible gates can be constructed in AML as we will see later.

In classical logic circuits, chaining \gate{and} and \gate{not} yields the universal gate \gate{nand} which can be used to implement arbitrary Boolean logic.
Similar concatenation can be performed in AFNs, provided care is taken to preserve the required ground states.
Non-trivial global effects mean that naively combining the networks for \gate{and} and \gate{not} in Fig.~\ref{fig:and_or_not} does not immediately yield \gate{nand}: upon merging the output edge for $X \land Y$ in Fig.~\ref{fig:and_or_not}a with the input edge in Fig.~\ref{fig:and_or_not}b, the new path from the $X$ input to the ground legs of the added \gate{not} means the resultant network has two ground states for the input combination $X=1$, $Y=0$, one with output $0$ and the other with output $1$.
Nevertheless, active networks can still be exploited to construct \gate{nand} in this fashion.
The naive concatenation fails because the configuration that should be sending zero input into the appended \gate{not} portion no longer has sufficient energy to retain its unique ground state given the added extra flow path.
Therefore, by taking the above construction and inserting an additional edge before what was the $X \lor Y$ output, the path from $X$ to this ignored output is made energetically preferable to the spurious new path from $X$ into the \gate{not} portion when $Y=0$. This restores the desired ground states (Fig.~\ref{fig:and_or_not}c).
Put broadly, if appending a new gate to an output, any additional candidate paths for input states intended to send zero into the appended portion can be disfavoured by upweighting all output and ground legs in the original portion through insertion of extra output-adjacent edges. In practice, such functional stabilization could be achievable by tuning channel geometry \cite{2016Wioland_RaceTracks}.

Another departure from classical circuits is in the \gate{fan-out} operation or signal splitter, taking one input and replicating it on two identical outputs.
Since active network flows are effectively discretized, a signal cannot be copied simply by splicing on another wire; rather, \gate{fan-out} is itself a powered and grounded active circuit.
In fact, since $X$ is output on both ground legs, \gate{fan-out} is realized simultaneously with \gate{not} in Fig.~\ref{fig:and_or_not}b, akin to the simultaneous realization of \gate{and} and \gate{or} in Fig.~\ref{fig:and_or_not}a.
This circuit can then be appended to any other gate to increase that gate's fan-out count, provided other output legs are lengthened as necessary in order to preserve the required ground state paths.
In general, as in traditional conservative logic~\cite{Fredkin82_IJTP}, the ground legs of an operation may output other useful logical expressions that can simplify construction of a larger system.
These considerations emphasize how AML is most effective as a top-down, global construction to benefit from the advantage of autonomy inherent in~AFNs.

\subsection*{Environmental noise}

Upon taking environmental noise into account, the Langevin equation \eqref{eq:langevin} holds and the density of states is a Boltzmann distribution $\propto \exp[-\beta H(\vPhi)]$.
AML then becomes probabilistic with a drop-off from perfect accuracy that can be tuned through geometry or microscopic activity \cite{2012Sanchez_Nature,2016Wioland_RaceTracks}. For $\beta \gg 1$ and $\mu \gg \lambda$, the order of magnitude of the error---the probability of observing the incorrect result---can be evaluated in terms of the free-edge flow weighting $\alpha = e^{-\beta \lambda / 12} \ll 1$ by replacing the continuous density of states $\exp[-\beta H(\phi)]$ with that of a discretized system $\vPhi \in \{-1,0,1\}^{|E|}$ and expanding probabilities in $\alpha$.
This estimation shows that while the error in the \gate{nand} gate in Fig.~\ref{fig:and_or_not}c is linear in $\alpha$, the \gate{and}/\gate{or} network in Fig.~\ref{fig:and_or_not}a is more robust for specific input combinations. In particular, on top of simple mass conservation demanding that \gate{and} and \gate{or} must be essentially perfect for identical inputs $X=Y$, \gate{and} exhibits error of only $O(\alpha^3)$ for $X=0$, $Y=1$.
This robustness---due to the difference in the number of active edges between the two possible output states---and that of the more sensitive $X=1$, $Y=0$ input pair can be further exponentially enhanced by lengthening the upper edge, trading simplicity for accuracy.
True operational error can be quantified by numerical evaluation of the marginal distribution $p(O_v | \vI)$ for the desired output vertex $v$, as shown for \gate{nand} in Fig.~\ref{fig:and_or_not}d by numerical integration of Eq.~\eqref{eq:langevin} at two noise amplitudes (Methods).
These results imply that the desired robustness and the noise characteristics of the realization scheme should be taken into account when designing AML systems.

\subsection*{Memory}

Multistable circuits with memory arise naturally within AML as dynamic networks with multiple ground states.
A classic set--reset (SR) latch requires at least two logic gates---two \gate{nand}s, for instance---and feedback loops between them.
In contrast, the global topological feedback inherent in AFNs mean that a memory circuit similar to the SR latch can be constructed very simply (Fig.~\ref{fig:SR_latch}).
When $S$ and $R$ are both $0$, the two network ground states correspond to outputs $Q = 0$ and $Q = 1$.
In the zero noise limit $\beta \rightarrow \infty$ these states are stable and the output will not change until one of $S$ or $R$ is changed.
On setting $S$ to $1$, the flow route from the power leg to ground is cut off and $Q$ immediately sets to $1$; upon releasing $S$, the stable state requiring fewest edge changes (and so nearest in state space) will be favored~\cite{2016Woodhouse_PNAS} and $Q = 1$ is set.
Conversely, setting $R$ to $1$ forces the power flow through the ground leg such that releasing $R$ then favors the state with $Q = 0$.
Implementing this circuit as a continuous active network obeying the Langevin equation~\eqref{eq:langevin}~\cite{2016Wioland_NPhys,2016Woodhouse_PNAS} confirms its memory properties at low noise (Fig.~\ref{fig:SR_latch}c; Methods).
Traditional SR latch behavior, where $Q = 0$ is output immediately on setting $R = 1$, would require a doubly-grounded network capable of dissipating both the power and reset signals.

\subsection*{Reversible gates}

The \gate{nand} gate in Fig.~\ref{fig:and_or_not}c is not reversible, since the $(X,Y) = (0,1)$ and $(1,0)$ inputs yield identical output and ground leg states: that is, the precise input state cannot be deduced from all readable output data.
However, employing closed loops within an AFN, exploiting mutual exclusivity of active flows, allows fully reversible gates to be constructed.
For example, a reversible \gate{xor} ($\oplus$) gate is provided by the 
two-output \gate{cnot} (controlled-\gate{not}) operation familiar to quantum logic~\cite{PhysRevLett.75.4714}, realized as the ground states of the network in Fig.~\ref{fig:controlled}a. This accepts a data input $X$ and a control input $C$, outputting $\lnot X$ if $C$ is $1$ and $X$ if $C$ is $0$, which is precisely $X \oplus C$, as well as always outputting $C$, giving a one-to-one mapping of input pairs $(X,C)$ to output pairs $(X \oplus C,C)$.
The more complex three-input Fredkin or \gate{cswap} gate~\cite{Fredkin82_IJTP}, which serves as the fundamental gate of conservative logic~\cite{Fredkin82_IJTP} and is remarkably both reversible and universal, can also be realized in AML (Fig.~\ref{fig:controlled}b). In general, mass conservation means that we expect fully reversible computing to be realizable within the AML framework.

\section*{Discussion}

Beyond logical operations, finding local energy minima of an AFN can be recast as a Boolean satisfiability problem (SAT) similar to those considered in DNA-based computing~\cite{1994Adleman_Science,1996Lipton_Science}. Given a Boolean formula $f(x_1,\ldots, x_n)$ with logical variables $x_j\in\{0,1\}$, the associated SAT problem asks whether one can find solutions of $f(x_1,\ldots, x_n)=1$. To connect this problem with AFNs, suppose that the underlying graph $\Gamma$ is closed---that is, it has no inputs or outputs ($\partial \Gamma = \emptyset$).
To each edge flux $\phi_e$ we associate a Boolean variable $x_e$ where $x_e = 1$ represents $\phi_e = \pm 1$ and $x_e = 0$ represents $\phi_e = 0$. Incompressibility at a vertex $v \in V$ then implies the logical condition that an even number of the incident edges $e \in E(v)$ have $x_e = 1$.
For example, a bulk degree-3 vertex with incident edge variables $x_1$, $x_2$, $x_3$ has incompressibility condition
\begin{align}
(\lnot x_1 \lor \lnot x_2 \lor \lnot x_3) \land (\lnot x_1 \lor x_2 \lor x_3) \land (x_1 \lor \lnot x_2 \lor x_3) \land (x_1 \lor x_2 \lor \lnot x_3) =1
\end{align}
in conjunctive normal form.
Combining vertices then yields a $k$-SAT problem---one posed as a large \gate{and} of $k$-term \gate{or}s---where $k$ is the maximum vertex degree in $\Gamma$, whose solutions are potential metastable states of $H$.
If $\Gamma$ contains no dioded (one-way) edges then these are all energy minima, with multiplicity determined by the number of orientations of the subgraph induced by those $e \in E$ with $x_e = 1$; if there are multiple dioded edges, then some logical states may not be orientable. Input and output vertices simply add further logical conditions: the former force edges to assume $1$ or $0$ values according to the input vector $\vI$, and the dioded edges of the latter reduce to fixing the number of $1$-valued output edges to be the number of inputs.
In general, many active biological network processes might be fruitfully cast in logical terms: \textit{P.\ polycephalum}, for instance, could be viewed as solving a constrained SAT problem to coarsen an initially fine foraging network~\cite{Tero2010_Science,Adamatzky_Book}.

The technology to implement AFN-based logic devices is now becoming available: electrostati\-cally-driven colloids~\cite{Bricard:2013aa,2016Yan_NMat}, self-propelled droplets~\cite{2011Shashi}, polar microfilaments~\cite{2010Schaller}, artificial extensile nematics~\cite{2012Sanchez_Nature} and microswimmer suspensions~\cite{2016Wioland_NPhys,2016Wioland_RaceTracks,2012Denissenko_PNAS,2015Paoluzzi_PRL} all present feasible AML realization schemes capable of sustaining microfluidic matter transport. A particularly promising candidate for the experimental implementation of AML schemes could be dense suspensions of \textit{Bacillus subtilis} bacteria, which spontaneously form robust unidirectional flows in loop-shaped microfluidic channels\cite{2016Wioland_RaceTracks}, provided the channel diameter is substantially smaller than the preferred bulk vortex size\cite{Wioland2013_PRL,Dunkel2013_PRL} ($\sim 70\,\mu$m) of an unconfined suspension. Such microbial suspensions are also amenable to rectification through microfluidic ratchets~\cite{Galajda2007,2012Denissenko_PNAS,2013Kantsler_PNAS}, thus enabling unidirectional `diode' edges. With regard to the future, the ideas developed here can be readily expanded to realize more complex design strategies by combining the basic AFNs discussed above with additional external control mechanisms such as optical activation~\cite{2013Palacci_Science} or chemical patterning~\cite{2016Uspal_PRL,Tompkins25032014}.

To conclude, AFNs present a flexible framework for biologically-rooted computing and autonomous lab-on-a-chip devices.
The non-local effects resulting from the interaction of active flows within complex topologies present interesting advantages over classical computing.
For instance, eavesdropping detection becomes near-trivial, since any snooping device installed within the network---or even on an ignored ground leg---is likely to fundamentally alter the ground states of the AFN, changing the output behavior and rendering the intrusion obvious.
Fully exploiting the global character of these systems to construct arbitrary computation will require innovative coupling of techniques from statistical physics, control theory and graph theory. Ultimately, this will lend insight into the natural optimization strategies that underly the balance between energy consumption and dissipation constraints present in biological systems.

\section*{Methods}

\begin{figure*}
\centering
\includegraphics{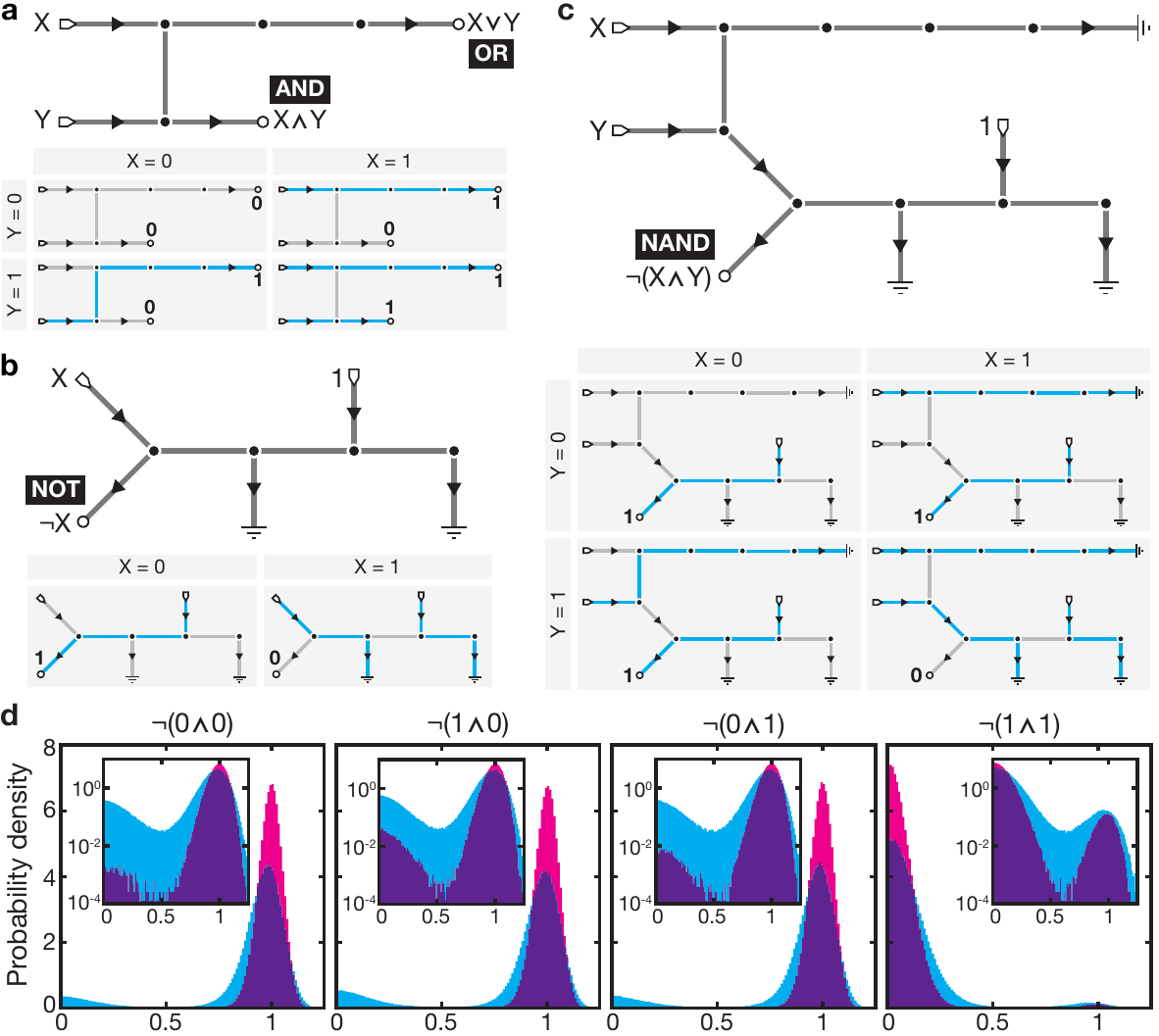}
\caption{
\onehalfspacing
\textbf{Elementary logical operations realized with AFNs} ---
\textbf{a}, An AFN whose ground states yield a simultaneous \gate{and}/\gate{or} gate. Either leg can be picked as the output depending on the operation desired, with the other sent to ground; alternatively, both could be used. Ground states for the four possible input combinations shown below, with active edges highlighted cyan.
\textbf{b}, Logical \gate{not} can be realized by a powered gate with two ground legs, as required by mass conservation.
\textbf{c}, The \gate{not} gate in \textbf{b} can be appended to the \gate{and} leg of \textbf{a} to yield \gate{nand}, provided the ground (\gate{or}) leg of \textbf{a} is lengthened to preserve the desired network states. 
\textbf{d}, Output histograms of the \gate{nand} gate in \textbf{c} at non-zero noise amplitudes, with $\beta\lambda = 100$ (magenta) and $\beta\lambda = 50$ (cyan); incompressibility is fixed at $\beta\mu = 500$ in both (Methods). Each histogram comprises $8\times 10^5$ data points. Inset: histograms with log-scaled vertical axis.
\label{fig:and_or_not}
}	
\end{figure*}

\begin{figure*}[h!]
\centering
\includegraphics{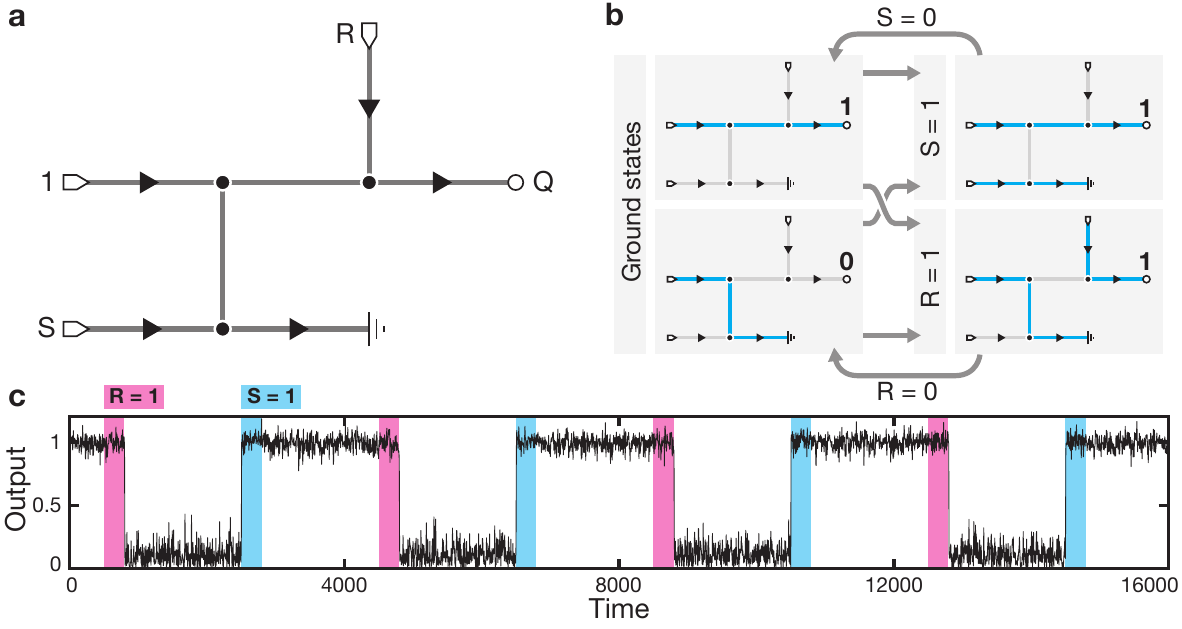}
\caption{
\onehalfspacing
\textbf{Intrinsic multi-stability of AFNs allows the simple construction of a 1-bit memory circuit} --- 
\textbf{a}, AFN for a circuit with SR latch-like behavior. $S$ and $R$ are set and reset inputs, respectively, used to control the network output $Q$.
\textbf{b}, With $S = R = 0$, the network has two ground states corresponding to $Q = 0$ and $Q = 1$. Raising to $S = 1$ forces an output of $1$, which is maintained when $S$ is released with high probability.
Conversely, pulsing $R = 1$ forces the system into the output $0$ ground state after $R$ is released. Mass conservation means that $Q = 1$ while $R = 1$; traditional SR latch behavior could be achieved with an additional ground leg at the expense of network complexity.
\textbf{c}, With low but non-zero noise, simulation of the network Langevin dynamics at $\beta\lambda = 100$, $\beta\mu = 500$ demonstrates the robust set--reset behavior of the network (Methods).
\label{fig:SR_latch}}
\end{figure*}

\paragraph*{Numerical integration}
Simulations displayed in Fig.~\ref{fig:and_or_not}d and Fig.~\ref{fig:SR_latch}c were performed by numerically integrating the Langevin equation \eqref{eq:langevin} using the Euler--Maruyama method~\cite{Higham2001_SIAMRev}. Diode (one-way) edges $e \in E_+$ were enforced with reflective boundary conditions at $\phi_e = 0$ by setting $\phi_e \rightarrow |\phi_e|$ at every integration step. Fig.~\ref{fig:and_or_not}d uses time step $\delta t = 2.5\times 10^{-3}$ and includes every 500th point up to $t = 10^6$, and Fig.~\ref{fig:SR_latch}c uses time step $\delta t = 5\times 10^{-3}$ and plots every 1000th point. Full Matlab code to perform these simulations is provided as Supplementary Software.

\begin{figure*}[h!]
\centering
\includegraphics{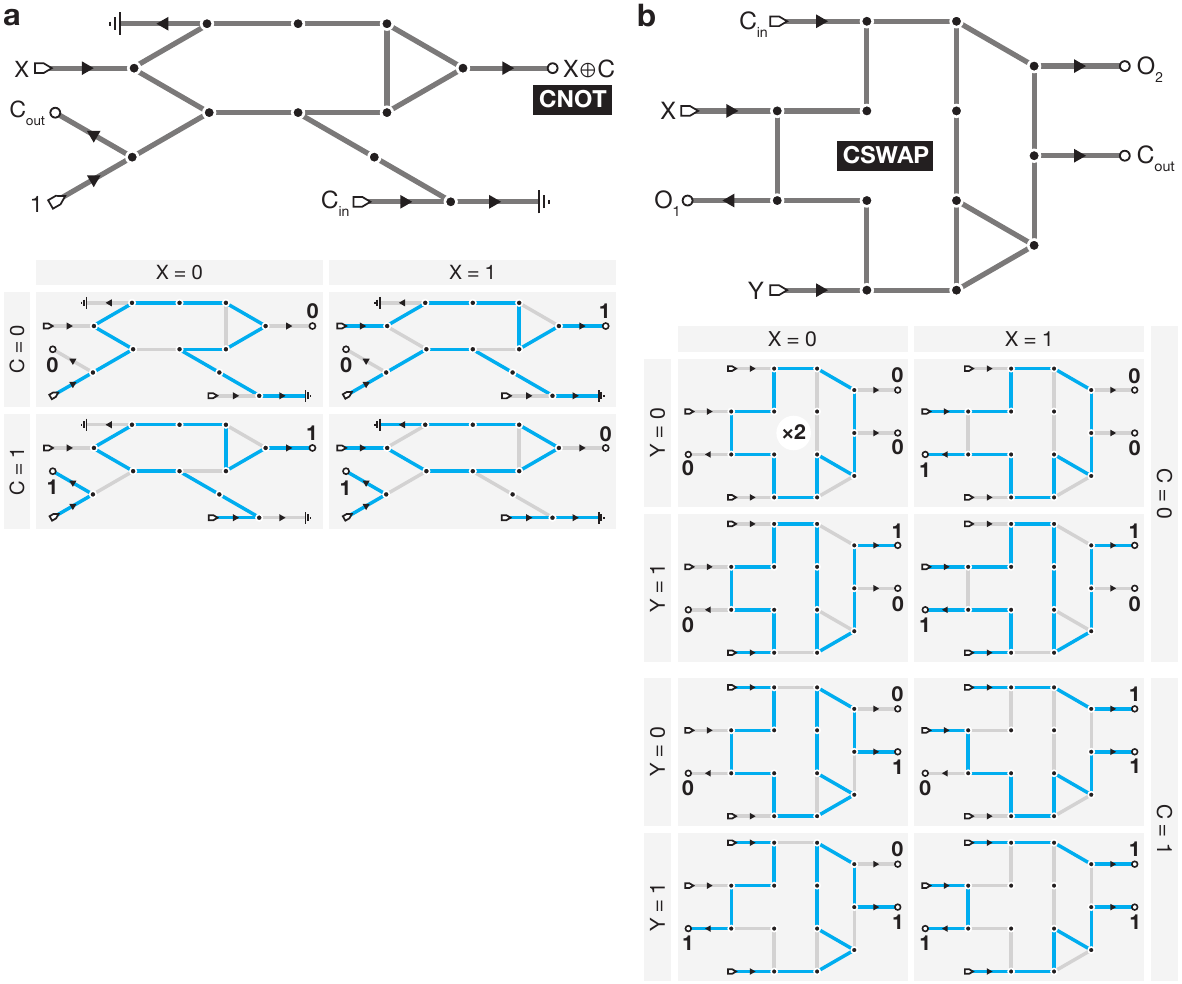}
\caption{
\onehalfspacing
\textbf{Reversible gates can be realized through AFNs employing cycles in their topology}~---
\textbf{a},~A~\gate{cnot} gate, which provides a reversible \gate{xor}, and its ground states. Though $C_\text{out}$ always equals $C_\text{in}$, it is not one flow path connecting the two. Instead, flow interactions mean that $C_\text{out}$ is drawn from the power leg.
\textbf{b}, A Fredkin gate, or \gate{cswap}, and its ground states. If $C_\text{in} = 0$, then $O_1 = X$ and $O_2 = Y$; if $C_\text{in} = 1$, then $O_1 = Y$ and $O_2 = X$. Since AML is conservative, this gate needs no power or ground legs. Note that the all-zero input state has two ground states corresponding to clockwise and counterclockwise orientations of the flow loop. The gate implementations in \textbf{a} and \textbf{b} were determined by an exhaustive numerical ground state search over constrained random graphs (Methods).
\label{fig:controlled}}	
\end{figure*}

\paragraph*{Network determination}
Networks in Figs.~\ref{fig:and_or_not} and \ref{fig:SR_latch} possessing ground states obeying the desired logical operations were determined analytically.
The \gate{cnot} and \gate{cswap} gates in Fig.~\ref{fig:controlled} were found by random search implemented in Mathematica. On generating a random graph $\Gamma$ with a degree distribution $\mathcal{D}$, comprising chosen fixed numbers of degree-$2$ and $3$ vertices and as many degree-$1$ vertices as total inputs and outputs, its ground states for all input combinations were determined by brute force evaluation over all discrete flows $\vPhi \in \{-1,0,1\}^{|E|}$. $\Gamma$ was then deemed a viable candidate if its ground states' outputs were unique for each input combination and these outputs followed the desired truth table. The numbers of degree-$2$ and degree-$3$ vertices in $\mathcal{D}$ were varied by trial and error until candidate graphs were found, typically taking on the order of thousands of random graph generations.
The chosen candidates were refined by hand to remove unnecessary complexity, and ground states were re-evaluated analytically and further checked against brute-force numerical calculation.

\paragraph*{Data availability}
Numerical integration code is provided in Supplementary Software.
Any further data available on request to the corresponding author.

\singlespacing

{
\small
\paragraph*{Acknowledgements}
This work was supported by Trinity College, Cambridge~(F.G.W.), an Alfred P. Sloan Research Fellowship~(J.D.), an Edmund F. Kelly Research Award~(J.D.), NSF Award CBET-1510768 (J.D.), and a Complex Systems Scholar Award of the James S. McDonnell Foundation (J.D.).

\paragraph*{Author contributions}
Both authors contributed at all stages of this work.

\paragraph*{Competing interests}
The authors declare no competing financial interests.
}

\clearpage

\begin{thebibliography}{10}
\expandafter\ifx\csname url\endcsname\relax
  \def\url#1{\texttt{#1}}\fi
\expandafter\ifx\csname urlprefix\endcsname\relax\def\urlprefix{URL }\fi
\providecommand{\bibinfo}[2]{#2}
\providecommand{\eprint}[2][]{\url{#2}}

\bibitem{2013Ge}
\bibinfo{author}{Ge, Q.}, \bibinfo{author}{Qi, H.~J.} \& \bibinfo{author}{Dunn,
  M.~L.}
\newblock \bibinfo{title}{Active materials by four-dimension printing}.
\newblock \emph{\bibinfo{journal}{Appl. Phys. Lett.}}
  \textbf{\bibinfo{volume}{103}}, \bibinfo{pages}{131901} (\bibinfo{year}{2013}).

\bibitem{2013Marchetti_Review}
\bibinfo{author}{Marchetti, M.~C.} \emph{et~al.}
\newblock \bibinfo{title}{Hydrodynamics of soft active matter}.
\newblock \emph{\bibinfo{journal}{Rev. Mod. Phys.}}
  \textbf{\bibinfo{volume}{85}}, \bibinfo{pages}{1143} (\bibinfo{year}{2013}).

\bibitem{Din:2016aa}
\bibinfo{author}{Din, M.~O.} \emph{et~al.}
\newblock \bibinfo{title}{Synchronized cycles of bacterial lysis for in vivo
  delivery}.
\newblock \emph{\bibinfo{journal}{Nature}} \textbf{\bibinfo{volume}{536}},
  \bibinfo{pages}{81--85} (\bibinfo{year}{2016}).

\bibitem{Tompkins25032014}
\bibinfo{author}{Tompkins, N.} \emph{et~al.}
\newblock \bibinfo{title}{Testing {T}uring's theory of morphogenesis in
  chemical cells}.
\newblock \emph{\bibinfo{journal}{Proc. Natl. Acad. Sci. U.S.A.}}
  \textbf{\bibinfo{volume}{111}}, \bibinfo{pages}{4397--4402}
  (\bibinfo{year}{2014}).

\bibitem{2014Karzbrun}
\bibinfo{author}{Karzbrun, E.}, \bibinfo{author}{Tayar, A.~M.},
  \bibinfo{author}{Noireaux, V.} \& \bibinfo{author}{Bar-Ziv, R.~H.}
\newblock \bibinfo{title}{Programmable on-chip {DNA} compartments as artificial
  cells}.
\newblock \emph{\bibinfo{journal}{Science}} \textbf{\bibinfo{volume}{345}},
  \bibinfo{pages}{829--832} (\bibinfo{year}{2014}).

\bibitem{1993Langer}
\bibinfo{author}{Langer, R.} \& \bibinfo{author}{Vacanti, J.~P.}
\newblock \bibinfo{title}{Tissue engineering}.
\newblock \emph{\bibinfo{journal}{Science}} \textbf{\bibinfo{volume}{260}},
  \bibinfo{pages}{920--926} (\bibinfo{year}{1993}).

\bibitem{di2010bacterial}
\bibinfo{author}{{Di Leonardo}, R.} \emph{et~al.}
\newblock \bibinfo{title}{Bacterial ratchet motors}.
\newblock \emph{\bibinfo{journal}{Proc. Natl. Acad. Sci. U.S.A.}}
  \textbf{\bibinfo{volume}{107}}, \bibinfo{pages}{9541--9545}
  (\bibinfo{year}{2010}).

\bibitem{Thampie1501854}
\bibinfo{author}{Thampi, S.~P.}, \bibinfo{author}{Doostmohammadi, A.},
  \bibinfo{author}{Shendruk, T.~N.}, \bibinfo{author}{Golestanian, R.} \&
  \bibinfo{author}{Yeomans, J.~M.}
\newblock \bibinfo{title}{Active micromachines: Microfluidics powered by
  mesoscale turbulence}.
\newblock \emph{\bibinfo{journal}{Sci. Adv.}} \textbf{\bibinfo{volume}{2}},
  \bibinfo{pages}{e1501854} (\bibinfo{year}{2016}).

\bibitem{2012Sanchez_Nature}
\bibinfo{author}{Sanchez, T.}, \bibinfo{author}{Chen, D. T.~N.},
  \bibinfo{author}{DeCamp, S.~J.}, \bibinfo{author}{Heymann, M.} \&
  \bibinfo{author}{Dogic, Z.}
\newblock \bibinfo{title}{Spontaneous motion in hierarchically assembled active
  matter}.
\newblock \emph{\bibinfo{journal}{Nature}} \textbf{\bibinfo{volume}{491}},
  \bibinfo{pages}{431--434} (\bibinfo{year}{2012}).

\bibitem{2014Keber_Science}
\bibinfo{author}{Keber, F.~C.} \emph{et~al.}
\newblock \bibinfo{title}{Topology and dynamics of active nematic vesicles}.
\newblock \emph{\bibinfo{journal}{Science}} \textbf{\bibinfo{volume}{345}},
  \bibinfo{pages}{1135--1139} (\bibinfo{year}{2014}).

\bibitem{2011Shashi}
\bibinfo{author}{Thutupalli, S.}, \bibinfo{author}{Seemann, R.} \&
  \bibinfo{author}{Herminghaus, S.}
\newblock \bibinfo{title}{Swarming behavior of simple model squirmers}.
\newblock \emph{\bibinfo{journal}{New J. Phys.}} \textbf{\bibinfo{volume}{13}},
  \bibinfo{pages}{073021} (\bibinfo{year}{2011}).

\bibitem{2008Walther_SM}
\bibinfo{author}{Walther, A.} \& \bibinfo{author}{Muller, A. H.~E.}
\newblock \bibinfo{title}{Janus particles}.
\newblock \emph{\bibinfo{journal}{Soft Matter}} \textbf{\bibinfo{volume}{4}},
  \bibinfo{pages}{663--668} (\bibinfo{year}{2008}).

\bibitem{2016Yan_NMat}
\bibinfo{author}{Yan, J.} \emph{et~al.}
\newblock \bibinfo{title}{Reconfiguring active particles by electrostatic
  imbalance}.
\newblock \emph{\bibinfo{journal}{Nat. Mater.}} \textbf{\bibinfo{volume}{15}},
  \bibinfo{pages}{1095--1099} (\bibinfo{year}{2016}).

\bibitem{2016DiLeonardo_NMat}
\bibinfo{author}{Di~Leonardo, R.}
\newblock \bibinfo{title}{Active colloids: {C}ontrolled collective motions}.
\newblock \emph{\bibinfo{journal}{Nat. Mater.}} \textbf{\bibinfo{volume}{15}},
  \bibinfo{pages}{1057--1058} (\bibinfo{year}{2016}).

\bibitem{Bricard:2013aa}
\bibinfo{author}{Bricard, A.}, \bibinfo{author}{Caussin, J.-B.},
  \bibinfo{author}{Desreumaux, N.}, \bibinfo{author}{Dauchot, O.} \&
  \bibinfo{author}{Bartolo, D.}
\newblock \bibinfo{title}{Emergence of macroscopic directed motion in
  populations of motile colloids}.
\newblock \emph{\bibinfo{journal}{Nature}} \textbf{\bibinfo{volume}{503}},
  \bibinfo{pages}{95--98} (\bibinfo{year}{2013}).

\bibitem{2012Sokolov}
\bibinfo{author}{Sokolov, A.} \& \bibinfo{author}{Aranson, I.~S.}
\newblock \bibinfo{title}{Physical properties of collective motion in
  suspensions of bacteria}.
\newblock \emph{\bibinfo{journal}{Phys. Rev. Lett.}}
  \textbf{\bibinfo{volume}{109}}, \bibinfo{pages}{248109}
  (\bibinfo{year}{2012}).

\bibitem{2016Wioland_NPhys}
\bibinfo{author}{Wioland, H.}, \bibinfo{author}{Woodhouse, F.~G.},
  \bibinfo{author}{Dunkel, J.} \& \bibinfo{author}{Goldstein, R.~E.}
\newblock \bibinfo{title}{Ferromagnetic and antiferromagnetic order in
  bacterial vortex lattices}.
\newblock \emph{\bibinfo{journal}{Nat. Phys.}} \textbf{\bibinfo{volume}{12}},
  \bibinfo{pages}{341--345} (\bibinfo{year}{2016}).

\bibitem{Snezhko:2011aa}
\bibinfo{author}{Snezhko, A.} \& \bibinfo{author}{Aranson, I.~S.}
\newblock \bibinfo{title}{Magnetic manipulation of self-assembled colloidal
  asters}.
\newblock \emph{\bibinfo{journal}{Nat. Mater.}} \textbf{\bibinfo{volume}{10}},
  \bibinfo{pages}{698--703} (\bibinfo{year}{2011}).

\bibitem{2016Wioland_RaceTracks}
\bibinfo{author}{Wioland, H.}, \bibinfo{author}{Lushi, E.} \&
  \bibinfo{author}{Goldstein, R.~E.}
\newblock \bibinfo{title}{Directed collective motion of bacteria under channel
  confinement}.
\newblock \emph{\bibinfo{journal}{New J. Phys.}} \textbf{\bibinfo{volume}{18}},
  \bibinfo{pages}{075002} (\bibinfo{year}{2016}).

\bibitem{Sokolov19012010}
\bibinfo{author}{Sokolov, A.}, \bibinfo{author}{Apodaca, M.~M.},
  \bibinfo{author}{Grzybowski, B.~A.} \& \bibinfo{author}{Aranson, I.~S.}
\newblock \bibinfo{title}{Swimming bacteria power microscopic gears}.
\newblock \emph{\bibinfo{journal}{Proc. Natl. Acad. Sci. U.S.A.}}
  \textbf{\bibinfo{volume}{107}}, \bibinfo{pages}{969--974}
  (\bibinfo{year}{2010}).

\bibitem{2015Pearce}
\bibinfo{author}{Pearce, D. J.~G.} \& \bibinfo{author}{Turner, M.~S.}
\newblock \bibinfo{title}{Emergent behavioural phenotypes of swarming models
  revealed by mimicking a frustrated anti-ferromagnet}.
\newblock \emph{\bibinfo{journal}{J. R. Soc. Interface}}
  \textbf{\bibinfo{volume}{12}}, \bibinfo{pages}{20150520} (\bibinfo{year}{2015}).

\bibitem{2016Nicolau_PNAS}
\bibinfo{author}{Nicolau~Jr., D.~V.} \emph{et~al.}
\newblock \bibinfo{title}{Parallel computation with molecular-motor-propelled
  agents in nanofabricated networks}.
\newblock \emph{\bibinfo{journal}{Proc. Natl. Acad. Sci. U.S.A.}}
  \textbf{\bibinfo{volume}{113}}, \bibinfo{pages}{2591--2596}
  (\bibinfo{year}{2016}).

\bibitem{2016Wang_EPJST}
\bibinfo{author}{Wang, A.~L.} \emph{et~al.}
\newblock \bibinfo{title}{Configurable {NOR} gate arrays from
  {B}elousov-{Z}habotinsky micro-droplets}.
\newblock \emph{\bibinfo{journal}{Eur. Phys. J. Spec. Top.}}
  \textbf{\bibinfo{volume}{225}}, \bibinfo{pages}{211--227}
  (\bibinfo{year}{2016}).

\bibitem{Tero2010_Science}
\bibinfo{author}{Tero, A.} \emph{et~al.}
\newblock \bibinfo{title}{Rules for biologically inspired adaptive network
  design}.
\newblock \emph{\bibinfo{journal}{Science}} \textbf{\bibinfo{volume}{327}},
  \bibinfo{pages}{439--442} (\bibinfo{year}{2010}).

\bibitem{Adamatzky_Book}
\bibinfo{author}{Adamatzky, A.}
\newblock \emph{\bibinfo{title}{Physarum Machines: Computers from Slime Molds}}
  (\bibinfo{publisher}{World Scientific}, \bibinfo{address}{Singapore},
  \bibinfo{year}{2010}).

\bibitem{Prakash832}
\bibinfo{author}{Prakash, M.} \& \bibinfo{author}{Gershenfeld, N.}
\newblock \bibinfo{title}{Microfluidic bubble logic}.
\newblock \emph{\bibinfo{journal}{Science}} \textbf{\bibinfo{volume}{315}},
  \bibinfo{pages}{832--835} (\bibinfo{year}{2007}).

\bibitem{Fuerstman828}
\bibinfo{author}{Fuerstman, M.~J.}, \bibinfo{author}{Garstecki, P.} \&
  \bibinfo{author}{Whitesides, G.~M.}
\newblock \bibinfo{title}{Coding/decoding and reversibility of droplet trains
  in microfluidic networks}.
\newblock \emph{\bibinfo{journal}{Science}} \textbf{\bibinfo{volume}{315}},
  \bibinfo{pages}{828--832} (\bibinfo{year}{2007}).

\bibitem{2015Katsikis_NPhy}
\bibinfo{author}{Katsikis, G.}, \bibinfo{author}{Cybulski, J.~S.} \&
  \bibinfo{author}{Prakash, M.}
\newblock \bibinfo{title}{Synchronous universal droplet logic and control}.
\newblock \emph{\bibinfo{journal}{Nat. Phys.}} \textbf{\bibinfo{volume}{11}},
  \bibinfo{pages}{588--596} (\bibinfo{year}{2015}).

\bibitem{2016Woodhouse_PNAS}
\bibinfo{author}{Woodhouse, F.~G.}, \bibinfo{author}{Forrow, A.},
  \bibinfo{author}{Fawcett, J.~B.} \& \bibinfo{author}{Dunkel, J.}
\newblock \bibinfo{title}{Stochastic cycle selection in active flow networks}.
\newblock \emph{\bibinfo{journal}{Proc. Natl. Acad. Sci. U.S.A.}}
  \textbf{\bibinfo{volume}{113}}, \bibinfo{pages}{8200--8205}
  (\bibinfo{year}{2016}).

\bibitem{Sipser_Book}
\bibinfo{author}{Sipser, M.}
\newblock \emph{\bibinfo{title}{Introduction to the Theory of Computation}}
  (\bibinfo{publisher}{Cengage Learning}, \bibinfo{address}{Boston, USA},
  \bibinfo{year}{2013}), \bibinfo{edition}{3rd} edn.

\bibitem{1961Landauer}
\bibinfo{author}{Landauer, R.}
\newblock \bibinfo{title}{Irreversibility and heat generation in the computing
  process}.
\newblock \emph{\bibinfo{journal}{IBM J. Res. Develop.}}
  \textbf{\bibinfo{volume}{5}}, \bibinfo{pages}{183--191}
  (\bibinfo{year}{1961}).

\bibitem{2012Berut_Nature}
\bibinfo{author}{Berut, A.} \emph{et~al.}
\newblock \bibinfo{title}{Experimental verification of {L}andauer's principle
  linking information and thermodynamics}.
\newblock \emph{\bibinfo{journal}{Nature}} \textbf{\bibinfo{volume}{483}},
  \bibinfo{pages}{187--189} (\bibinfo{year}{2012}).

\bibitem{PhysRevLett.75.4714}
\bibinfo{author}{Monroe, C.}, \bibinfo{author}{Meekhof, D.~M.},
  \bibinfo{author}{King, B.~E.}, \bibinfo{author}{Itano, W.~M.} \&
  \bibinfo{author}{Wineland, D.~J.}
\newblock \bibinfo{title}{Demonstration of a fundamental quantum logic gate}.
\newblock \emph{\bibinfo{journal}{Phys. Rev. Lett.}}
  \textbf{\bibinfo{volume}{75}}, \bibinfo{pages}{4714--4717}
  (\bibinfo{year}{1995}).

\bibitem{1986Hopfield_Science}
\bibinfo{author}{Hopfield, J.~J.} \& \bibinfo{author}{Tank, D.~W.}
\newblock \bibinfo{title}{Computing with neural circuits: A model}.
\newblock \emph{\bibinfo{journal}{Science}} \textbf{\bibinfo{volume}{223}},
  \bibinfo{pages}{625--633} (\bibinfo{year}{1986}).

\bibitem{Lestas:2010aa}
\bibinfo{author}{Lestas, I.}, \bibinfo{author}{Vinnicombe, G.} \&
  \bibinfo{author}{Paulsson, J.}
\newblock \bibinfo{title}{Fundamental limits on the suppression of molecular
  fluctuations}.
\newblock \emph{\bibinfo{journal}{Nature}} \textbf{\bibinfo{volume}{467}},
  \bibinfo{pages}{174--178} (\bibinfo{year}{2010}).

\bibitem{1994Adleman_Science}
\bibinfo{author}{Adleman, L.~M.}
\newblock \bibinfo{title}{Molecular computation of solutions to combinatorial
  problems}.
\newblock \emph{\bibinfo{journal}{Science}} \textbf{\bibinfo{volume}{266}},
  \bibinfo{pages}{1021--1024} (\bibinfo{year}{1994}).

\bibitem{1996Lipton_Science}
\bibinfo{author}{Lipton, R.~J.}
\newblock \bibinfo{title}{{DNA} solution of hard computational problems}.
\newblock \emph{\bibinfo{journal}{Science}} \textbf{\bibinfo{volume}{268}},
  \bibinfo{pages}{542--545} (\bibinfo{year}{1996}).

\bibitem{Hopfield_PNAS}
\bibinfo{author}{Hopfield, J.~J.}
\newblock \bibinfo{title}{Neural networks and physical systems with emergent
  collective computational abilities}.
\newblock \emph{\bibinfo{journal}{Proc. Natl. Acad. Sci. U.S.A.}}
  \textbf{\bibinfo{volume}{79}}, \bibinfo{pages}{2554--2558}
  (\bibinfo{year}{1982}).

\bibitem{2012Denissenko_PNAS}
\bibinfo{author}{Denissenko, P.}, \bibinfo{author}{Kantsler, V.},
  \bibinfo{author}{Smith, D.~J.} \& \bibinfo{author}{Kirkman-Brown, J.}
\newblock \bibinfo{title}{Human spermatozoa migration in microchannels reveals
  boundary-following navigation}.
\newblock \emph{\bibinfo{journal}{Proc. Natl. Acad. Sci. U.S.A.}}
  \textbf{\bibinfo{volume}{109}}, \bibinfo{pages}{8007--8010}
  (\bibinfo{year}{2012}).

\bibitem{1995Toner_PRL}
\bibinfo{author}{Toner, J.} \& \bibinfo{author}{Tu, Y.}
\newblock \bibinfo{title}{Long-range order in a two-dimensional dynamical {XY}
  model: {H}ow birds fly together}.
\newblock \emph{\bibinfo{journal}{Phys. Rev. Lett.}}
  \textbf{\bibinfo{volume}{75}}, \bibinfo{pages}{4326} (\bibinfo{year}{1995}).

\bibitem{1998Toner_PRE}
\bibinfo{author}{Toner, J.} \& \bibinfo{author}{Tu, Y.}
\newblock \bibinfo{title}{Flocks, herds, and schools: A quantitative theory of
  flocking}.
\newblock \emph{\bibinfo{journal}{Phys. Rev. E}} \textbf{\bibinfo{volume}{58}},
  \bibinfo{pages}{4828} (\bibinfo{year}{1998}).

\bibitem{2006Schneidman}
\bibinfo{author}{Schneidman, E.}, \bibinfo{author}{Berry, M.~J.},
  \bibinfo{author}{Segev, R.} \& \bibinfo{author}{Bialek, W.}
\newblock \bibinfo{title}{Weak pairwise correlations imply strongly correlated
  network states in a neural population}.
\newblock \emph{\bibinfo{journal}{Nature}} \textbf{\bibinfo{volume}{440}},
  \bibinfo{pages}{1007--1012} (\bibinfo{year}{2006}).

\bibitem{2015Chen_NJP}
\bibinfo{author}{Chen, L.}, \bibinfo{author}{Toner, J.} \&
  \bibinfo{author}{Lee, C.~F.}
\newblock \bibinfo{title}{Critical phenomenon of the order--disorder transition
  in incompressible active fluids}.
\newblock \emph{\bibinfo{journal}{New J. Phys.}} \textbf{\bibinfo{volume}{17}},
  \bibinfo{pages}{042002} (\bibinfo{year}{2015}).

\bibitem{2016Chen_NatCommun}
\bibinfo{author}{Chen, L.}, \bibinfo{author}{Lee, C.~F.} \&
  \bibinfo{author}{Toner, J.}
\newblock \bibinfo{title}{Mapping two-dimensional polar active fluids to
  two-dimensional soap and one-dimensional sandblasting}.
\newblock \emph{\bibinfo{journal}{Nat. Commun.}} \textbf{\bibinfo{volume}{7}},
  \bibinfo{pages}{12215} (\bibinfo{year}{2016}).

\bibitem{Souslov2016}
\bibinfo{author}{Souslov, A.}, \bibinfo{author}{{van Zuiden}, B.~C.},
  \bibinfo{author}{Bartolo, D.} \& \bibinfo{author}{Vitelli, V.}
\newblock \bibinfo{title}{Topological sound in active-liquid metamaterials}.
\newblock Preprint at \emph{\bibinfo{journal}{http://arxiv.org/abs/1610.06873v2}}  (\bibinfo{year}{2016}).

\bibitem{Fredkin82_IJTP}
\bibinfo{author}{Fredkin, E.} \& \bibinfo{author}{Toffoli, T.}
\newblock \bibinfo{title}{Conservative logic}.
\newblock \emph{\bibinfo{journal}{Int. J. Theor. Phys.}}
  \textbf{\bibinfo{volume}{21}}, \bibinfo{pages}{219--253}
  (\bibinfo{year}{1982}).

\bibitem{2010Schaller}
\bibinfo{author}{Schaller, V.}, \bibinfo{author}{Weber, C.},
  \bibinfo{author}{Semmrich, C.}, \bibinfo{author}{Frey, E.} \&
  \bibinfo{author}{Bausch, A.~R.}
\newblock \bibinfo{title}{Polar patterns of driven filaments}.
\newblock \emph{\bibinfo{journal}{Nature}} \textbf{\bibinfo{volume}{467}},
  \bibinfo{pages}{73--77} (\bibinfo{year}{2010}).

\bibitem{2015Paoluzzi_PRL}
\bibinfo{author}{Paoluzzi, M.}, \bibinfo{author}{Di~Leonardo, R.} \&
  \bibinfo{author}{Angelani, L.}
\newblock \bibinfo{title}{Self-sustained density oscillations of swimming
  bacteria confined in microchambers}.
\newblock \emph{\bibinfo{journal}{Phys. Rev. Lett.}}
  \textbf{\bibinfo{volume}{115}}, \bibinfo{pages}{188303}
  (\bibinfo{year}{2015}).

\bibitem{Wioland2013_PRL}
\bibinfo{author}{Wioland, H.}, \bibinfo{author}{Woodhouse, F.~G.},
  \bibinfo{author}{Dunkel, J.}, \bibinfo{author}{Kessler, J.~O.} \&
  \bibinfo{author}{Goldstein, R.~E.}
\newblock \bibinfo{title}{Confinement stabilizes a bacterial suspension into a
  spiral vortex}.
\newblock \emph{\bibinfo{journal}{Phys. Rev. Lett.}}
  \textbf{\bibinfo{volume}{110}}, \bibinfo{pages}{268102}
  (\bibinfo{year}{2013}).

\bibitem{Dunkel2013_PRL}
\bibinfo{author}{Dunkel, J.} \emph{et~al.}
\newblock \bibinfo{title}{Fluid dynamics of bacterial turbulence}.
\newblock \emph{\bibinfo{journal}{Phys. Rev. Lett.}}
  \textbf{\bibinfo{volume}{110}}, \bibinfo{pages}{228102}
  (\bibinfo{year}{2013}).

\bibitem{Galajda2007}
\bibinfo{author}{Galajda, P.}, \bibinfo{author}{Keymer, J.},
  \bibinfo{author}{Chaikin, P.} \& \bibinfo{author}{Austin, R.}
\newblock \bibinfo{title}{{A wall of funnels concentrates swimming bacteria.}}
\newblock \emph{\bibinfo{journal}{J. Bacteriol.}}
  \textbf{\bibinfo{volume}{189}}, \bibinfo{pages}{8704--8707}
  (\bibinfo{year}{2007}).

\bibitem{2013Kantsler_PNAS}
\bibinfo{author}{Kantsler, V.}, \bibinfo{author}{Dunkel, J.},
  \bibinfo{author}{Polin, M.} \& \bibinfo{author}{Goldstein, R.~E.}
\newblock \bibinfo{title}{Ciliary contact interactions dominate surface
  scattering of swimming eukaryotes}.
\newblock \emph{\bibinfo{journal}{Proc. Natl. Acad. Sci. USA}}
  \textbf{\bibinfo{volume}{110}}, \bibinfo{pages}{1187--1192}
  (\bibinfo{year}{2013}).

\bibitem{2013Palacci_Science}
\bibinfo{author}{Palacci, J.}, \bibinfo{author}{Sacanna, S.},
  \bibinfo{author}{Steinberg, A.~P.}, \bibinfo{author}{Pine, D.~J.} \&
  \bibinfo{author}{Chaikin, P.~M.}
\newblock \bibinfo{title}{Living crystals of light-activated colloidal
  surfers}.
\newblock \emph{\bibinfo{journal}{Science}} \textbf{\bibinfo{volume}{339}},
  \bibinfo{pages}{936--940} (\bibinfo{year}{2013}).

\bibitem{2016Uspal_PRL}
\bibinfo{author}{Uspal, W.~E.}, \bibinfo{author}{Popescu, M.~N.},
  \bibinfo{author}{Dietrich, S.} \& \bibinfo{author}{Tasinkevych, M.}
\newblock \bibinfo{title}{Guiding catalytically active particles with
  chemically patterned surfaces}.
\newblock \emph{\bibinfo{journal}{Phys. Rev. Lett.}}
  \textbf{\bibinfo{volume}{117}}, \bibinfo{pages}{048002}
  (\bibinfo{year}{2016}).

\bibitem{Higham2001_SIAMRev}
\bibinfo{author}{Higham, D.~J.}
\newblock \bibinfo{title}{An algorithmic introduction to numerical simulation
  of stochastic differential equations}.
\newblock \emph{\bibinfo{journal}{SIAM Rev.}} \textbf{\bibinfo{volume}{43}},
  \bibinfo{pages}{525--546} (\bibinfo{year}{2001}).

\end{thebibliography}
\end{document}